\documentclass[%
 reprint,
 amsmath,amssymb,
 aps,
]{revtex4-1}

\usepackage{graphicx}
\usepackage{dcolumn}
\usepackage{bm}

\usepackage{tabularx}
\usepackage{xcolor}

\begin{document}

\preprint{APS/123-QED}

\title{Coexistence of vector soliton Kerr combs in normal dispersion resonators}
\author{B. Kostet$^{a}$, Y. Soupart$^{a}$, K. Panajotov$^{b,c}$, and M. Tlidi$^{a}$}
\affiliation{$^{a}$Département de Physique, Faculté des Sciences, Université Libre
de Bruxelles (U.L.B.), CP 231, Campus Plaine, B-1050 Bruxelles, Belgium}
\affiliation{$^{b}$Department of Applied Physics and Photonics (IR-TONA), Vrije Universiteit Brussels, Pleinlaan 2, 1050 Brussels, Belgium}
\affiliation{$^{c}$Institute of Solid State Physics, 72 Tzarigradsko Chaussee Blvd., 1784 Sofia, Bulgaria}

\begin{abstract}
We investigate the formation of dark vector localized structures in the presence of nonlinear polarization mode coupling in optical resonators subject to a coherent optical injection in the normal dispersion regime.  This simple device is described by coupled Lugiato-Lefever equations.  The stabilization of localized structures is attributed to a front locking mechanism. We show that in a multistable homogeneous steady-state regime, two branches of dark localized structures can coexist for a fixed value of the system parameters. These coexisting solutions possess different polarization states and different power peaks in the microresonator. We characterize in-depth their formation by drawing their bifurcation diagrams in regimes close to modulational instability and far from it. It is shown that both branches of localized structures exhibit a heteroclinic collapse snaking type of behavior. The coexistence of two vectorial branches of dark localized states is not possible without taking into account polarization degrees of freedom. 
\end{abstract}

\maketitle


\section{Introduction}
Localized structures (LSs) often called dissipative solitons is a patterning phenomenon that has been widely encountered in many far from equilibrium systems including fluid mechanics, optics, biology, and medicine (see overviews on this issue \cite{akhmediev2008dissipative,chembo2017theory,tlidi2018dissipative1,tlidi2018dissipative2,malomed2019nonlinear}). These coherent and robust structures are often characterized by an intrinsic wavelength that is solely determined by the dynamical parameters and not by system size or boundary conditions.\\ A classic example of nonequilibrium systems that undergo instabilities leading to the formation of LSs that are experimentally accessible is the field of nonlinear optics and laser physics. In particular, in broad area optical resonators where light suffers diffraction, LSs consist of one or more spots of light in a two-dimensional transverse plane embedded on a homogeneous background. However, in small area devices, by using a waveguide in both transverse directions, the intra-cavity field can be spatially stabilized. In this case, diffraction can be neglected in the modeling, and replaced by a natural chromatic dispersion due to the Kerr medium which affects both the amplitude and the phase of light circulating within the cavity. The advantage of taking dispersion into account instead of diffraction is that chromatic dispersion varies with the wavelength and can be tuned. LSs in the temporal domain can thus be formed and their optical spectra consist of equidistant lines forming frequency combs. The link between LSs and frequency comb generation has been established allowing to reinforce the interest in the field of LS formation (see an excellent review by Lugiato et al.  \cite{lugiato2018lugiato} in Theme issues \cite{tlidi2018dissipative1,tlidi2018dissipative2}).\\ Optical frequency combs generated by cavity resonators possess a wide spectrum of applications in science and technology, ranging from high-precision spectroscopy to metrology and photonics~\cite{fortier201920}. Increasing interest has been paid to dissipative soliton frequency combs which correspond, in the time domain, to stable temporal localized structures or pulses that propagate within the cavity with the group velocity of light. They were reported experimentally in microresonators~\cite{delhaye_optical_2007,Herr2014}.

\begin{figure}
		\includegraphics[width=0.42\linewidth]{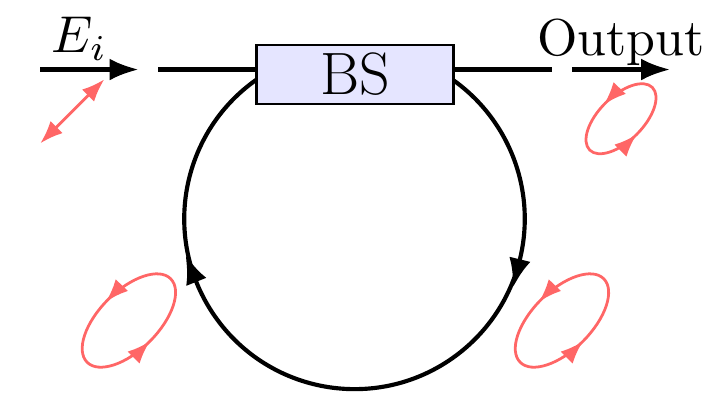}
	\includegraphics[width=0.56\linewidth]{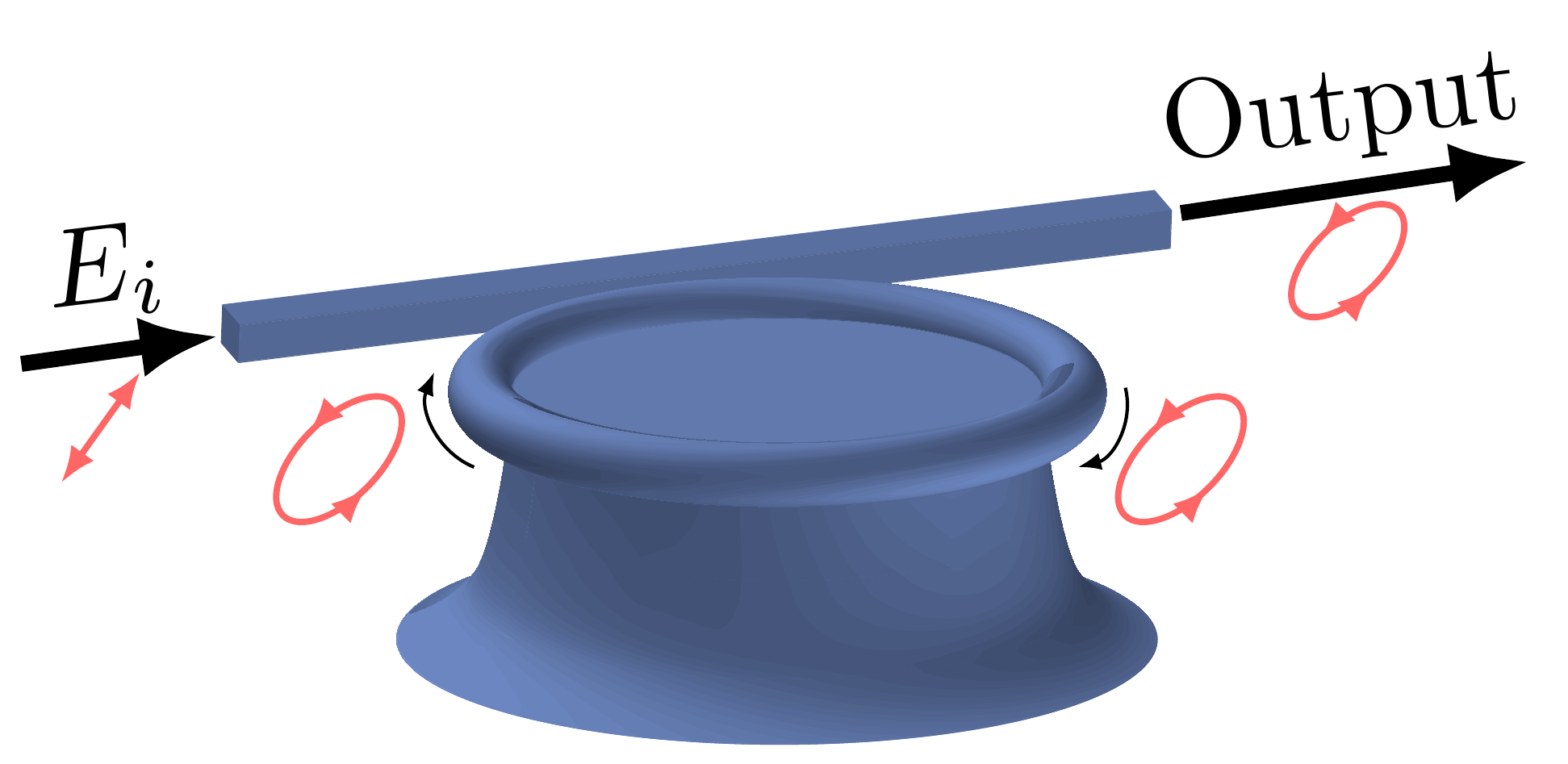}\\
			\caption{Schematics of two typical devices that can be described by the Lugiato-Lefever model: (left) macroscopic fiber resonator (right) microscopic toroidal resonator. Both devices are injected with linearly polarized light. }
	\label{fig:LLE}
\end{figure}

When taking into account polarization degree of freedom, it has been shown that during the propagation in Kerr media, new modulational instabilities can appear~\cite{PhysRevA.38.2018}, leading to polarization domain wall vector solitons~\cite{Haelterman:94bis}, symmetry breaking~\cite{PhysRevLett.79.661,PhysRevLett.89.083901} and soliton bound states~\cite{Kockaert:99}.\\ In particular, polarization properties of bright dissipative solitons frequency combs  have been the subject of recent investigation both theoretically~\cite{averlant_coexistence_2017,PhysRevLett.122.013905,8588302,saha_polarization_2020} and experimentally~\cite{PhysRevLett.126.023904}. Vector dissipative solitons (VDSs) are classified in two types:
Polarization-Locked Vector Solitons (PL-VS) and Group-Velocity-Locked Vector Solitons (GVL-VS). The former have been first predicted~\cite{PhysRevE.49.5742} and experimentally verified~\cite{PhysRevLett.82.3988} while
GVL-VS have been investigated in dispersion-managed cavity fiber laser. VDSs however do not consist solely of these two most common cases, but can also take the form of, e.g.\ polarization precessing Vector
Locked Solitons as demonstrated in the spatial domain~\cite{doi:10.1098/rsta.2014.0006,Mou:11,Mou:13}.

The paper is organized as follows. After an introduction, the well-known coupled Lugiato-Lefever equations describing driven microcavity by taking into account polarization degrees of freedom is introduced in Sec.~\ref{sec:2}.  We provide a detailed linear stability analysis by drawing a map as a function of the injection strength and the detuning parameter. In Sec.~\ref{sec:3}, we discuss the mechanism of front locking leading to the formation of localized structures in the normal dispersion regime, and we study the bifurcation structure in a multistable homogeneous steady-state regime. In Sec.~\ref{sec:4} we analyze the situation where the two branches of dark localized structures exhibit an overlapping domain of stability. In this case, we identify a wide range of parameters in which we observe the coexistence of dark localized structures with different polarization states and different power peaks. We conclude in Sec.~\ref{sec:5}.

\section{A coupled Lugiato-Lefever model and its linear stability analysis}\label{sec:2}

We consider nonlinear optical resonators filled with a Kerr medium. Macro- and micro-resonators are schematically depicted in Fig.~\ref{fig:LLE}. A continuous wave with components along the tranverse x and y axes $E_{i_x}$ and $E_{i_y}$ launched into the cavity through a beam splitter, propagates inside the Kerr medium, and experiences dispersion and
the Kerr effect. At each round trip, the light inside the
fiber is coherently superimposed with the input
beam. This can be described by the boundary conditions and the propagation equations consist of two-coupled nonlinear Schrodinger equations. The formulation of this problem leads to infinite-dimensional maps that can be simplified to a coupled LLE equations by applying the mean-field approximation for macroresonators such as all-fiber cavities ~\cite{Haelterman:94} and for microresonators~\cite{Hansson:18}. The dimensionless coupled LLE model reads 
\begin{eqnarray}
\label{eq:LLV}
\partial_{t}E_{x,y} &=& E_{i_x,i_y}+ i\left(|E_{x,y}|^2 + b |E_{y,x}|^2\right)E_{x,y}\\ \nonumber
&-&\left(1+i\theta_{x,y}\mp \Delta\beta_{1}\partial_{\tau}-i\beta_2 \partial_{\tau\tau}\right) E_{x,y}.
\end{eqnarray}
$E_{x,y}$ are the slowly varying envelopes for each polarization component of the intracavity field. 
We will focus on the case where the linearly polarized injected field has the same intensity for each polarization component, $E_{i_x}=E_{i_y}=E_i$. $b$ is the cross-phase modulation coefficient between the two components. $\theta_{x}$ and $\theta_{y}$ are the frequency detunings for each polarization direction. $\Delta\beta_{1}$ corresponds to the group velocity mismatch (GVM), associated with the first-order dispersion, while the second-order dispersion coefficient $\beta_2$ is assumed to be the same in each direction. In the present study, we will focus on a macroscopic fiber resonator cavity. This allows us to set the cross-phase modulation coefficient as $b=2/3$~\cite{AGRAWAL2013193} and to reasonably approximate the GVM $\Delta\beta_{1}$ as zero~\cite{6720193,PhysRevLett.123.013902}. As we set $b<1$, we operate in a regime where polarization domain walls do not occur~\cite{PhysRevLett.126.023904,Haelterman:94bis}. We also consider the normal dispersion regime, $\beta_2=-1$.

The steady-state continuous wave solutions of the coupled LLE (\ref{eq:LLV}) satisfy:
\begin{eqnarray}
\label{eqHSS}
I_{i_x,i_y}= \left[1+\left(\theta_{x,y}-I_{x,y} - b I_{y,x}\right)^2\right]I_{x,y}
\end{eqnarray}
where $I_{i_x,i_y}=E_{i_x,i_y}^2$ and $I_{x,y}=|E_{x,y}|^2$ are the intensities of  injected fields and intensities of the intracavity fields, respectively.   The Eqs.~(\ref{eqHSS})  possess up to nine solutions for fixed values of the system parameters, five of which are physical. In a previous communication~\cite{Kostet:21}, we have limited the analysis to parameter values for which the CW resembles the scalar case where the system develops the classical bistable S-curve. The linear stability analysis of the CW solutions with respect to a finite frequency of the form $\exp{(i\omega \tau+\lambda t)}$ is performed.\\
\begin{figure}
	\includegraphics[width=1.15\linewidth]{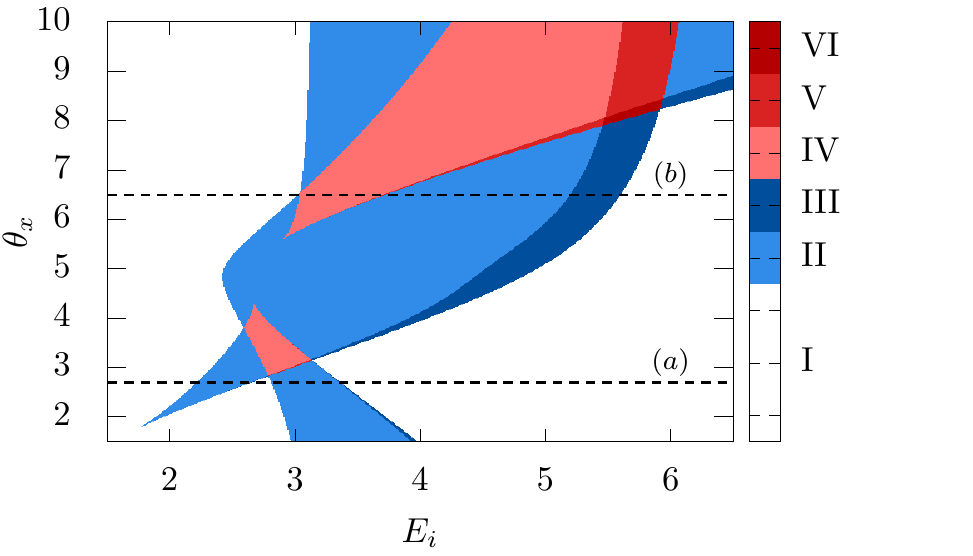}\\
	\vspace*{-0.4cm}
	\caption{Stability regions in the parameter space $E_i$--$\theta_x$. Parameters are: $\theta_y = 5$. Blue shades indicate regions of bistability, red shades indicate regions of tristability. Region I corresponds to monostability, with the presence of only one stable state. Region II corresponds to bistability between two stable states. Region III corresponds to bistability between one stable state and one modulationally unstable state. Region IV corresponds to tristability between three stable states. Region V corresponds to tristability between two stable states and one modulationally unstable state. Finally, region VI corresponds to tristability between one stable state and two modulationally unstable states. Examples of two consecutive bistable curves and of a tristable curve along the dashed lines (a) and (b) are shown in Fig.~\ref{fig:LSA}(a) and Fig.~\ref{fig:LSA}(b) respectively.}\vspace*{-0.5cm}
	\label{fig:map}
\end{figure}
The results are summarized in  the $E_i$--$\theta_x$ parameter space shown in Fig.~\ref{fig:map}. For small values of $\theta_x$, only a simple bistability exists, like in the scalar case as mentioned above (region II). As the detuning $\theta_x$ is increased, two well separated hysteresis loops, i.e., two separate regions of bistability appear (light blue regions in Fig.~\ref{fig:map}). Increasing further $\theta_x$ causes the two distinct bistabilities to get closer until they overlap. In this case, a tristability region is generated, in light red on the map (region IV). Proceeding along the map, this region collapses into the central bistability (II) zone. For even higher values of $\theta_x$, the tristability region (IV) reappears.
Every region is bordered from below by a domain of coexistence with one or two modulationally unstable states (III, V and VI). These domains with modulational instability sometimes get so thin that the spatial resolution of the plot does not allow to see them.\\
\begin{figure}
	\includegraphics[width=\linewidth]{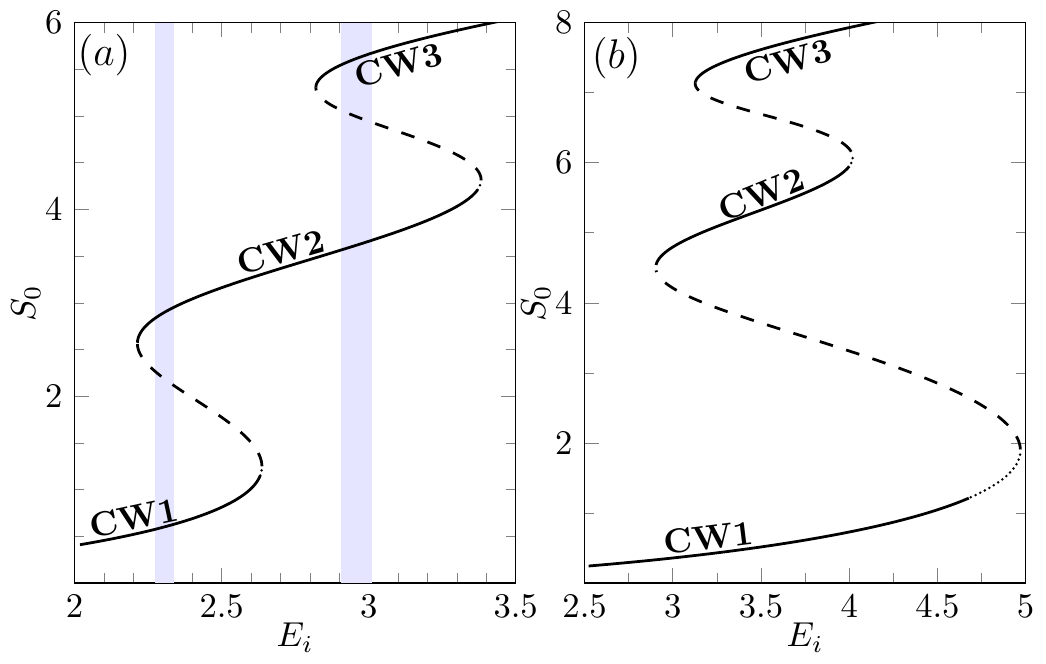}
	\caption{Homogeneous steady states. Two consecutive bistable curve Fig.~\ref{fig:LSA}(a) and tristable curve Fig.~\ref{fig:LSA}(b) taken along the dashed lines (a) and (b) in Fig.~\ref{fig:map} corresponding to $\theta_x = 2.7$ and $\theta_x = 6.5$, respectively. Full lines correspond to stable states, dashed lines correspond to unstable states and dotted lines correspond to modulationally unstable states. Highlighted regions in (a) correspond to zones where TLSs can be stabilized.}
	\label{fig:LSA}
\end{figure}
\begin{figure}
	\includegraphics[width=\linewidth]{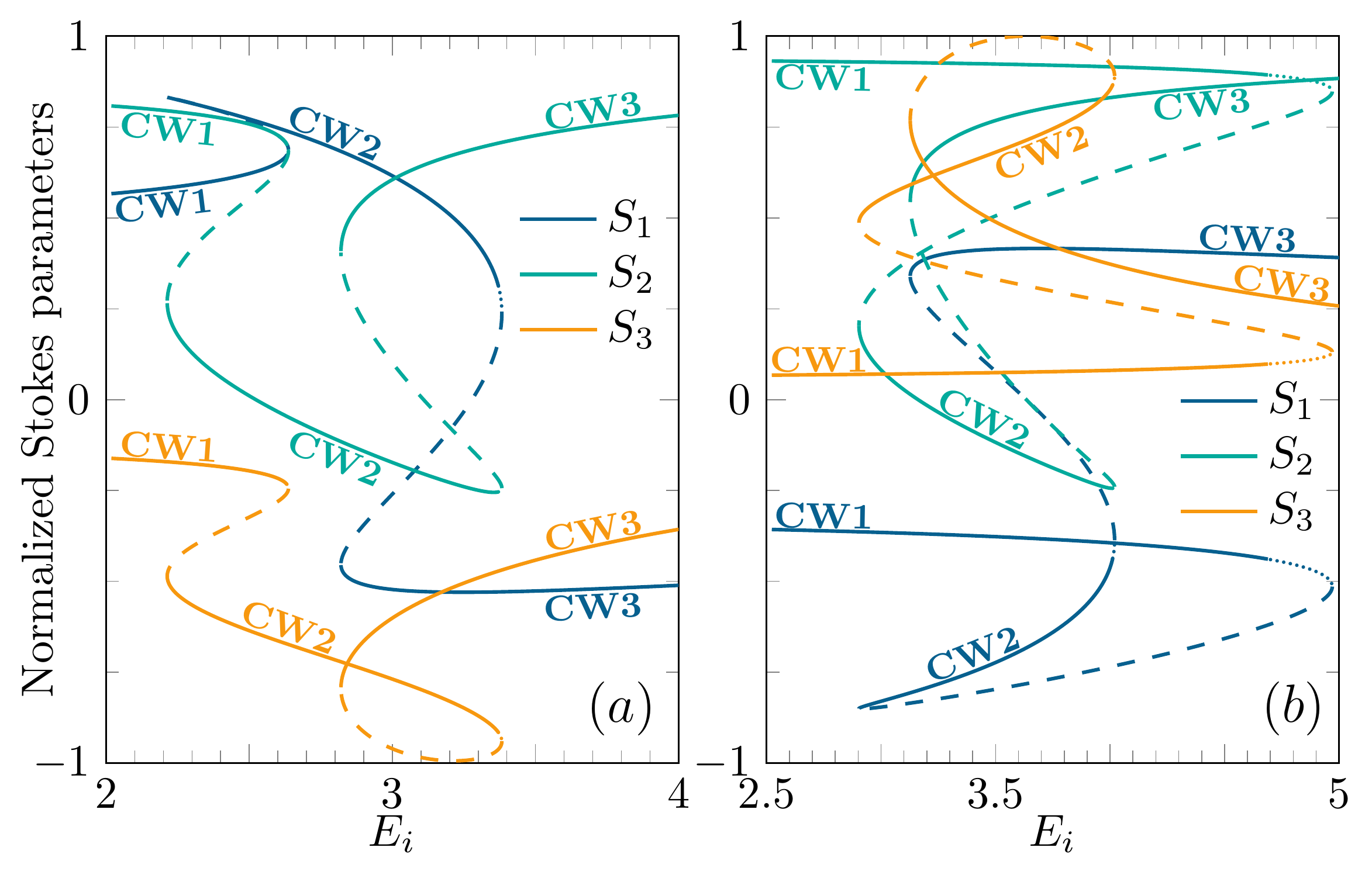}\\
	\caption{Normalized Stokes parameters. (a) CW solutions plotted in Fig.~\ref{fig:LSA}(a). (b) CW solutions plotted in Fig.~\ref{fig:LSA}(b).}
	\label{fig:HSSstokes}
\end{figure}
In what follows, we consider two examples of bistable and tristable curves Fig.~\ref{fig:LSA}(a) and Fig.~\ref{fig:LSA}(b), respectively. These figures are obtained from cuts along the dashed lines (a) and (b) in Fig.~\ref{fig:map}. The upper bistability collapses during the transition from region IV into region II in Fig.~\ref{fig:map}. This upper bistability reappears for higher values of the detuning parameter $\theta_x$. The small portions of modulational instability can be seen as dotted curves in Fig.~\ref{fig:LSA}(a) and \ref{fig:LSA}(b).\\

We first investigate the polarization properties of the homogeneous steady states shown in Fig.~\ref{fig:LSA}. The results are shown in Fig.~\ref{fig:HSSstokes} through the normalized Stokes parameters defined as $S_0=|E_x|^2+|E_y|^2$,  $S_1=(|E_x|^2-|E_y|^2)/S_0$,  $S_2=2\textrm{Re}(E_xE_y^*)/S_0$, $S_3=-2i(E_xE_y^*)/S_0$ (where ($^*$) stands for the complex conjugate). $S_0$ represents the total intensity of light in the cavity as seen above. $S_1$ corresponds to light polarized linearly along the axes x and y. $S_2$ also corresponds to light polarized linearly, but this time diagonally at 45 degrees with respect to the x and y axes. $S_3$ is indicative of the circularly polarized component of the light. In Fig.~\ref{fig:HSSstokes}(a), we start at low values of the injection with $S_1$ and $S_2$ significantly nonzero, while $S_3$ is close to zero, meaning that the light is mostly linearly polarized, with weak ellipticity.  As the injection is increased, all three Stokes parameters undergo an hysteresis loop, leading to a first bistability between CW$_1$ and CW$_2$ as in Fig.~\ref{fig:LSA}. In particular, $S_1$ increases to almost one in a very narrow hysteresis loop, $S_2$ decreases to almost zero, and $S_3$ increases strongly in absolute value, so that CW$_2$ gradually gets a strong ellipticity as the injection increases.
Finally, we have the second hysteresis loop leading to CW$_3$. For $S_1$ and $S_2$, the loops are quite large, ending on quite big values for both, with stronger $S_2$, than $S_1$. For $S_3$, the bistability starts from a high value and appears with the curve folding on itself, before getting closer to zero, reaching almost to the same values as in the beginning, so that CW$_3$ will also be mostly linearly polarized, with ellipticity  that decreases as the injection increases, but that stays higher than for CW$_1$.\\ In summary, both of the bistabilities from Fig.~\ref{fig:HSSstokes}(a) occur between states that have quite different polarization properties, especially regarding their ellipticity.\\ 
Following a similar process along the different curves in Fig.~\ref{fig:HSSstokes}(b), we can see that in this case, CW$_1$ has an almost completely diagonal linear polarization with $S_1$ being low, $S_2$ being almost one and $S_3$ being almost zero. The polarization properties of CW$_1$ barely change with the change in injection power. CW$_2$ has a very strong ellipticity due to $S_1$ keeping quite high absolute values, $S_2$ being close to zero and $S_3$ being in the same range of values as $S_1$. Finally, CW$_3$ has a constant $S_1$ at an intermediate value, while its $S_2$ increases to high values and $S_3$ gradually decreases to values close to zero as the injection power increases,  leading to a mostly diagonally linearly polarized light.\\ In addition to the same conclusion as for the previous case, we can notice by comparing  Fig.~\ref{fig:HSSstokes}(a) and (b) that, without changing the injection intensity or the polarization of the injected light, merely changing the value of the detuning along one of the axis led to very significant changes in the polarization properties of the solutions, as well as in the size and shape of the hysteresis loops corresponding to the Stokes parameters.

\section{Bifurcation structure for the dark localized states for double bistability:
Collapsed snaking}\label{sec:3}
\begin{figure*}
	\includegraphics[width=0.7\linewidth]{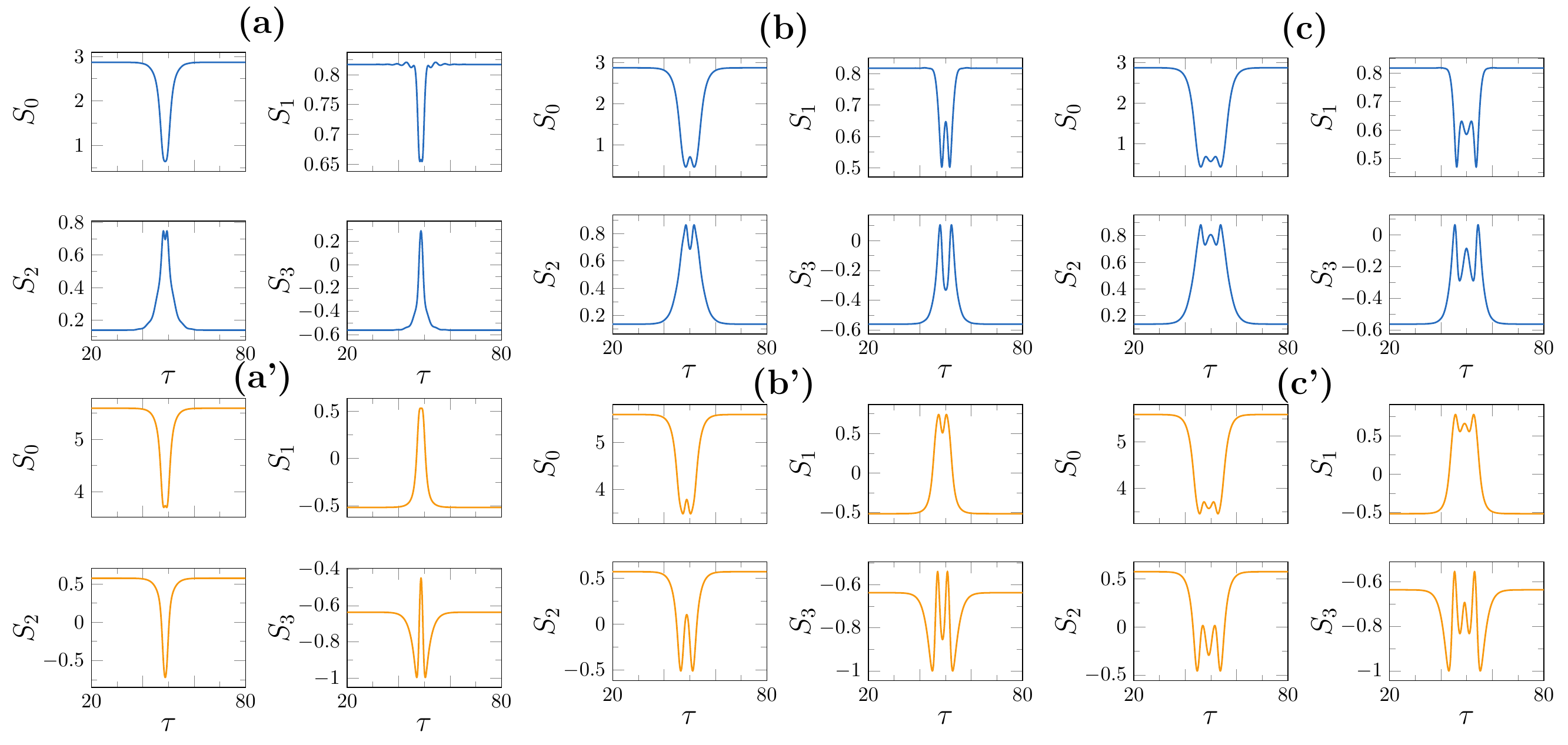}\\
	\caption{Profiles of the Stokes parameters $S_0$, $S_1$, $S_2$ and $S_3$ as a function of the fast time $\tau$ for the stable TLSs solutions created in the highlighted regions of Fig.~\ref{fig:LSA}(a). Profiles (a-c) correspond to the region highlighted on the left, while (a'-c') correspond to the region highlighted on the right. Injection power values are $E_i =$ \textrm{(a)} 2.2917 \textrm{(b)} 2.2922 \textrm{(c)} 2.2926 \textrm{(a')} 2.9394 \textrm{(b')} 2.9406 \textrm{(c')} 2.9406.}
	\label{fig:doublebistabstokes}
\end{figure*}
\begin{figure}
	\includegraphics[width=1\linewidth]{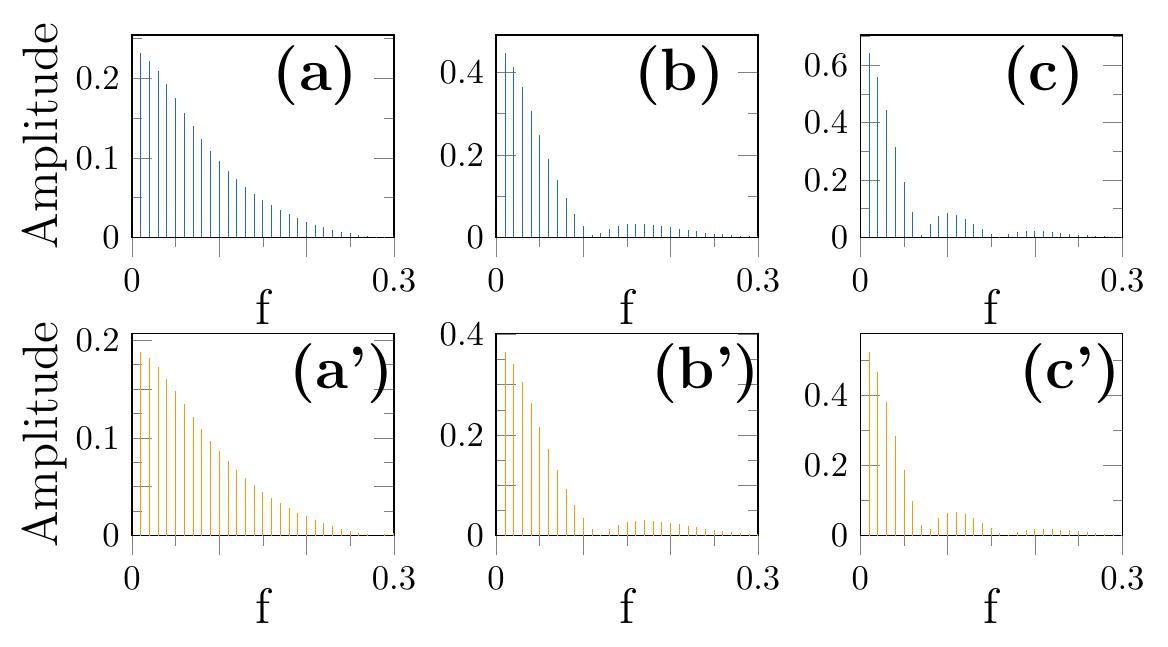}\\
	\caption{Vector Kerr combs corresponding to the stable solutions shown in Fig.~\ref{fig:doublebistabstokes}.}
	\label{fig:doublebistabspec}
\end{figure}
In dispersive Kerr micro- and macro-resonators, temporal localized structures (TLSs) can be generated in both anomalous and normal dispersion regimes. In the former case, TLSs appear thanks to the subcritical nature of the modulation instability and their formation does not require a commutation process between the CW solutions. They can be generated in the monostable regime.  \cite{scroggie_pattern_1994,tlidi_localized_1994}. In the temporal regime, these TLSs exhibit homoclinic snaking bifurcation diagrams that has been established first close to zero dispersion wavelength \cite{tlidi_high-order_2010}. In this case, dark TLS exhibit multistability behavior in a finite range of parameters referred to as pinning region where their bifurcation diagram consists of two snaking curves: one describes TLSs with $2n$ dips, while the other corresponds to $2n + 1$ dips with n a positive integer. As one moves further along the snaking curve, the TLS becomes better localized and acquires stability at the turning point where the slope becomes infinite. Afterwards, the TLS begins to grow along the fast temporal $\tau$ axis by adding extra dips symmetrically at either side. This growth is associated with back and forth oscillations across the pinning interval. This behavior is referred to as homoclinic snaking, which have been established first in the spatial domain \cite{champneys_homoclinic_1998,woods_heteroclinic_1999}.

However, to generate dark TLSs in the normal dispersion regime, it is necessary that the optical resonators operate in the bistable regime. In this case the formation of dark TLS results from switching wave connecting the two stable CW solutions of the input-output characteristics curve \cite{coen_convection_1999,xue_mode-locked_2015,garbin_experimental_2017}. These solutions are then of heteroclinic nature and their bifurcation diagram exhibits collapsed snaking \cite{parra-rivas_dark_2016,parra-rivas_origin_2016}. More recently, this type of behavior has been reported in a vectorial driven all-fiber resonators \cite{Kostet:21} in a simple bistable regime. They appear when two fronts connecting stable CW states interact and lock into a stationary and robust temporal dark dissipative structure. In what follows, we focus on that type of solutions in a regime far from the MI instabilities, and we first consider the situation where the CW solutions develop a double bistability curve with well separated hysteresis loops as shown in the parameter range corresponding to the highlighted regions of Fig.~\ref{fig:LSA}(a).

In this regime, we find stable heteroclinic TLS solutions and we plot the profiles of their normalized Stokes parameters in Fig.~\ref{fig:doublebistabstokes}. All these profiles correspond to heteroclinic connections between the homogeneous steady states shown in Fig.~\ref{fig:LSA}(a) and Fig.~\ref{fig:HSSstokes}(a). For these TLS solutions, the background corresponds to CW$_2$ for the profiles of Fig.~\ref{fig:doublebistabstokes}(a-c) and to CW$_3$ for the profiles  of Fig.~\ref{fig:doublebistabstokes}(a'-c'). The connection for (a-c) is to CW$_1$ and to CW$_2$ for (a'-c'). Note that the profiles of the $S_3$ component shown in Fig.~\ref{fig:doublebistabstokes} being non-zero is indicative of the presence of an elliptical light polarization in the cavity. Furthermore, we can see that the different coexisting TLS solutions connecting the same two CW states exhibit different polarization properties.\\ The spectral content of the total intensity profiles $S_0$ are optical frequency combs. The combs corresponding to the profiles from Fig.~\ref{fig:doublebistabstokes} are shown in Fig.~\ref{fig:doublebistabspec}. The comb lines are all equally spaced since the free spectral range, given by the inverse of the cavity roundtrip time of their associated TLSs is always the same. Their envelopes depend however on the exact profile of the considered solution. As the TLSs become wider when moving down the snaking curve, the comb becomes narrower. For the fundamental solution, the envelope of the spectrum follows an expected sech$^2$ curve~\cite{Herr2014}, but for each new solution appearing at each SN bifurcation, the new bump emerging in the $S_0$ profile also brings a new bump in the envelope of the spectrum.\\

\begin{figure}
	\includegraphics[width=\linewidth]{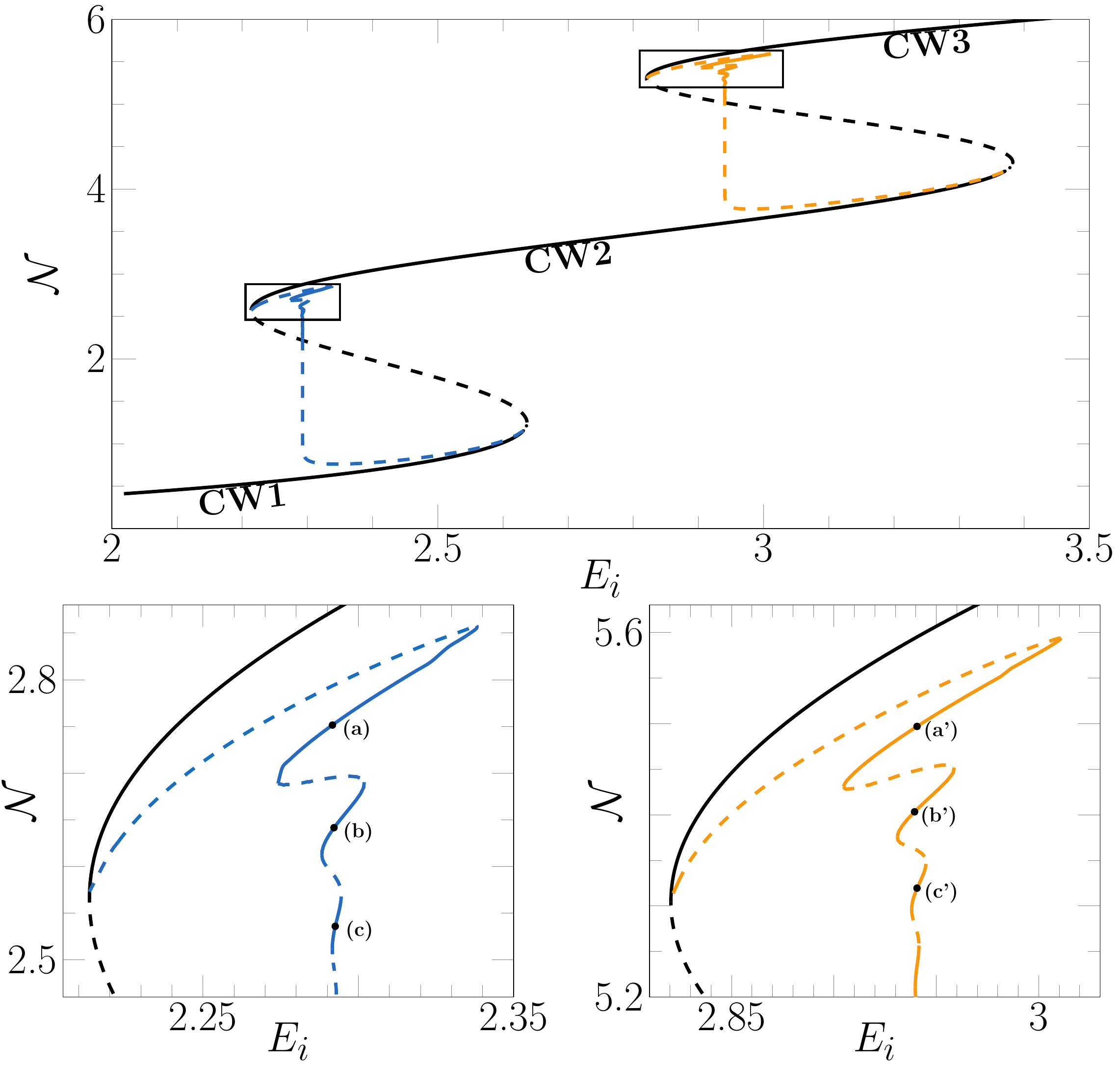}\\
	\caption{Double heteroclinic snaking. Top panel: Bifurcation diagram showing the L2-norm $\mathcal{N}$ as a function of the injected field intensity $E_i$. Stable (unstable) states are denoted by solid (dashed) lines. MI states are denoted by dotted lines. Parameters are: $\theta_x = 2.7, \theta_y = 5$. Bottom panels: Close-ups on the snaking curves collapsing onto the Maxwell point of each respective bistability.}
	\label{fig:doublebistab}
\end{figure}
\begin{figure}
	\includegraphics[width=0.75\linewidth]{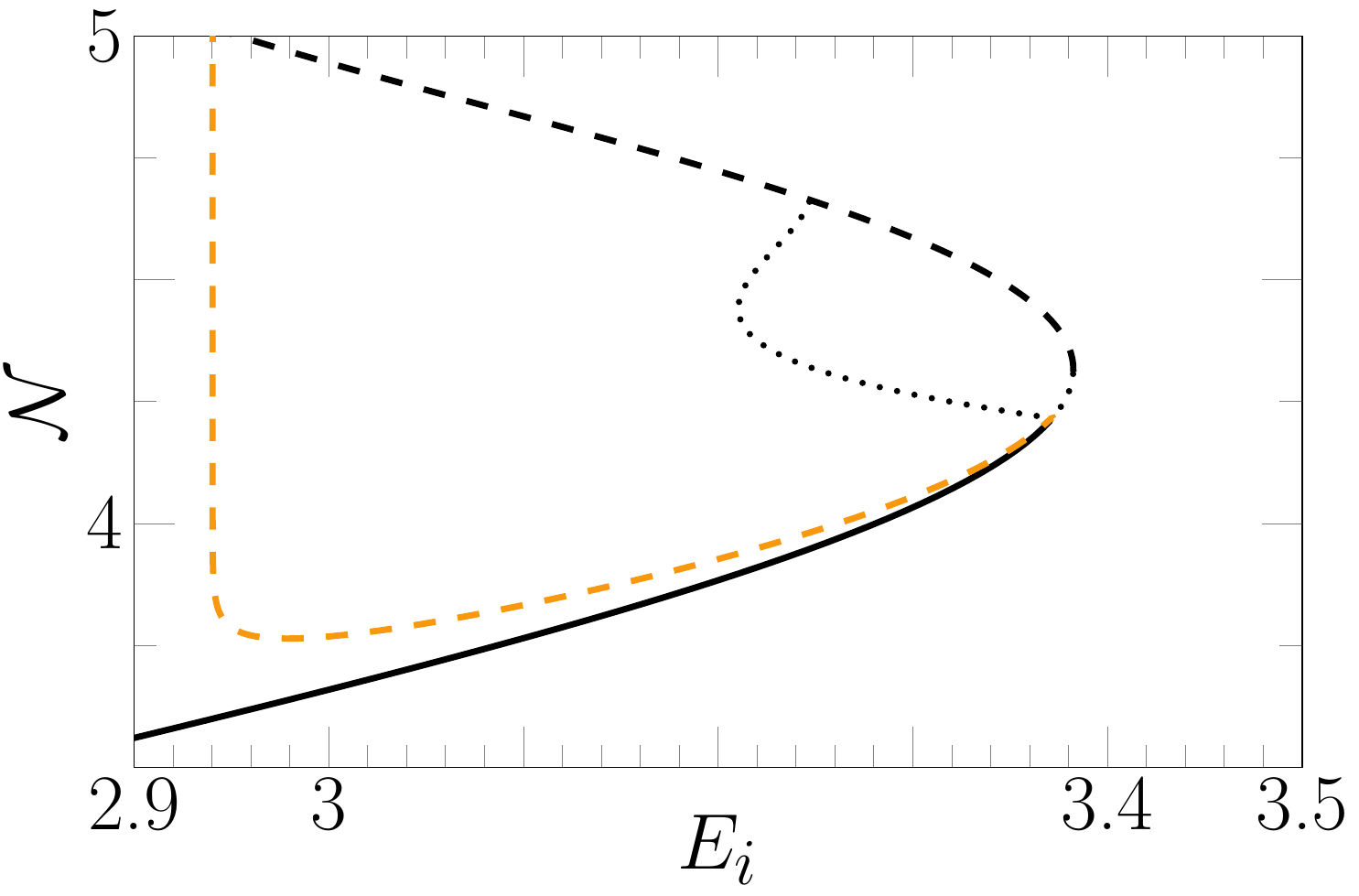}\\
	\caption{Close-up showing the MI branch emerging from the end of the upper snaking curve from Fig.~\ref{fig:doublebistab}.}
	\label{fig:MI}
\end{figure}

To construct the bifurcation diagram associated with vectorial TLS, we fix the detuning parameters and chromatic dispersion coefficient, and let the injected field amplitude be the control parameter.  We numerically simulate Eqs.\ref{eq:LLV} to initiate a predictor-corrector continuation method in the parameter space~\cite{10.1145/779359.779362}. Periodic boundary conditions are considered. The result is shown in the top panel of Fig.~\ref{fig:doublebistab}(a), where $\mathcal{N} = \int S_0/L\, \textrm{d}\tau$ is the normalized L2-norm of the electrical field with $L$ the size of the system. When increasing $E_i$, a branch of vectorial TLS indicated in dark blue emerges from the upper limit point or turning point associated with the bistability curve that involves the CW$_1$ and CW$_2$ solutions. In order to visualize better this first family of vectorial TLS, a zoom around the dark blue curve is shown in Fig.~\ref{fig:doublebistab}(b). This branch correspond to the profiles (a-c) shown in Fig.~\ref{fig:doublebistabstokes}.\\
As we increase the injection intensity $E_i$, a region of monostability for CW$_2$ arises, followed by another turning point leading to a second bistability between the CW$_2$ and CW$_3$ solutions, shown in light orange. A second distinct branch of vectorial TLS emerges from the upper turning point of this new bistability, as shown in the zoom shown in Fig.~\ref{fig:doublebistab}(c). This branch corresponds to the profiles (a'-c') shown in Fig.~\ref{fig:doublebistabstokes}.\\
These two branches undergo a collapsed snaking: they emerge from the upper saddle-node bifurcation of their respective bistabilities; then oscillate with exponentially damped amplitude before collapsing on the Maxwell point where the fronts are stationary~\cite{PhysRevLett.62.2957}. Finally, they end up connecting to the modulational instability bifurcation point on the lower part of their respective bistabilities. This MI branch emerges from this point, and connects to the unstable branch as shown in Fig~\ref{fig:MI}. This figure indicates that the range of parameters for which our TLSs exist is indeed far from the MI branch so that they will not be affected by it. \\

This shows that for this regime of parameters, two different branches, or families of TLS solutions coexist, but for different values of the injection field intensity, so that the two families cannot coexist in the same physical system. These TLS branches, during their oscillations, undergo several saddle-node (SN) bifurcations creating multiple different stable TLSs. At each of these bifurcations, the new stable TLS solution is characterized by the arising of a new bump at its bottom and becomes wider.\\
\begin{figure*}[t]
	\includegraphics[width=0.8\linewidth]{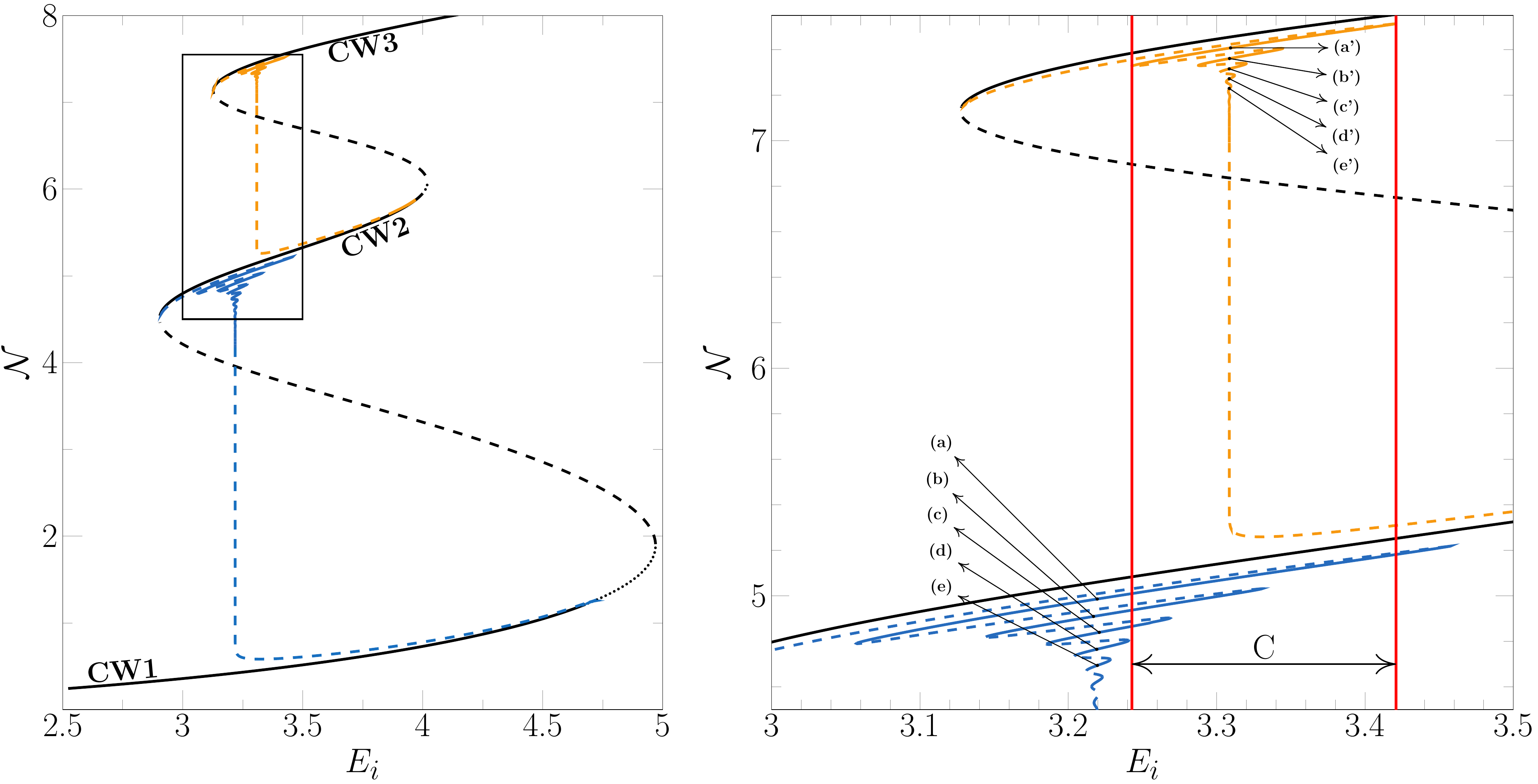}\\
	\caption{Tristable heteroclinic snaking. Left panel: Bifurcation diagram showing the normalized amount of photons $\mathcal{N}$ as a function of the injected field intensity $E_i$. Stable (unstable) states are denoted by solid (dashed) lines. MI states are denoted with dotted lines. Parameters are: $\theta_x = 6.5, \theta_y = 4.5$. Right panel: Close-up on the snaking curves collapsing onto the Maxwell point of each respective bistability showing the coexistence region $C$.}
	\label{fig:tristab}
\end{figure*}
\section{Coexistence between two vectorial branches of dark localized states}\label{sec:4}

In the scalar case, i.e., without taking into account the polarization degree of freedom, the third-order dispersion allows a coexistence between bright and dark TLS \cite{parra-rivas_coexistence_2017}. This behavior occurs far from any MI, and the resulting TLS connects the CWs of high and low ntensities. However, both of these solutions move since the third-order dispersion breaks the reflection symmetry. Recently, it has been shown that the LLE without high order dispersions, support moving TLSs \cite{clerc_time-delayed_2020,clerc2020nonlocal}. The motion of TLSs is attributed to the Raman delayed nonlocal response of the fiber. This effect can induce a coexistence between bright and dark TLSs \cite{parra2021influence}.

When considering polarization degree of freedom, a coexistence between two types of bright localized structures has been reported in a weakly birefringent all-fiber cavity subject to a linearly polarized optical injection, in the anomalous dispersion regime. The resulting vector TLSs do not only differ by their polarization states but also by their peak intensities ~\cite{averlant_coexistence_2017}. They are generated in a regime where two modulational instabilities with different frequencies appear in the system . Recently, it has been shown that in the normal dispersion regime, the collapsed snaking structure of the TLSs formed through switching waves interaction in the absence of modulational instability allows for the coexistence between multiple types of dark localized structures, which also differ by their polarization states and by their shapes~\cite{Kostet:21}. In that work, all coexisting TLS  solutions connect the same upper and lower CW states. In what follows, we show that two branches of dark TLSs connecting more than two CW states are possible due to the formation of tristability induced by polarization degree of freedom. For this values of system parameters, we have constructed the bifurcation diagram associated with TLSs in the same way as described above. The results are shown in Fig.~\ref{fig:tristab}. Once again, two branches of vectorial TLSs emerge from the upper turning point of the two different bistability curves linking each pair of CW states, before going through damped oscillations and collapsing onto the modulational instability point of the lower turning point. 
Figure~\ref{fig:tristab} illustrates the remarkable property of domain $C$, namely that it is a region where the system exhibits a high degree of multistability. Beside the tristability associated with CW solutions, which are both stable, two families of dark TLSs can be obtained for fixed values of the system's parameters. This behavior is caused by the two branches of heteroclinic collapsed snakings exhibiting an overlapping domain of stability, as indicated by region C. In this case, the two families are composed of the solutions (a-c) from the bistability connecting CW$_1$ and CW$_2$ and the solutions (a'-e') from the bistability connecting CW$_2$ and CW$_3$, shown in the right part of Fig.~\ref{fig:tristab}.
\begin{figure*}
	\includegraphics[width=0.75\linewidth]{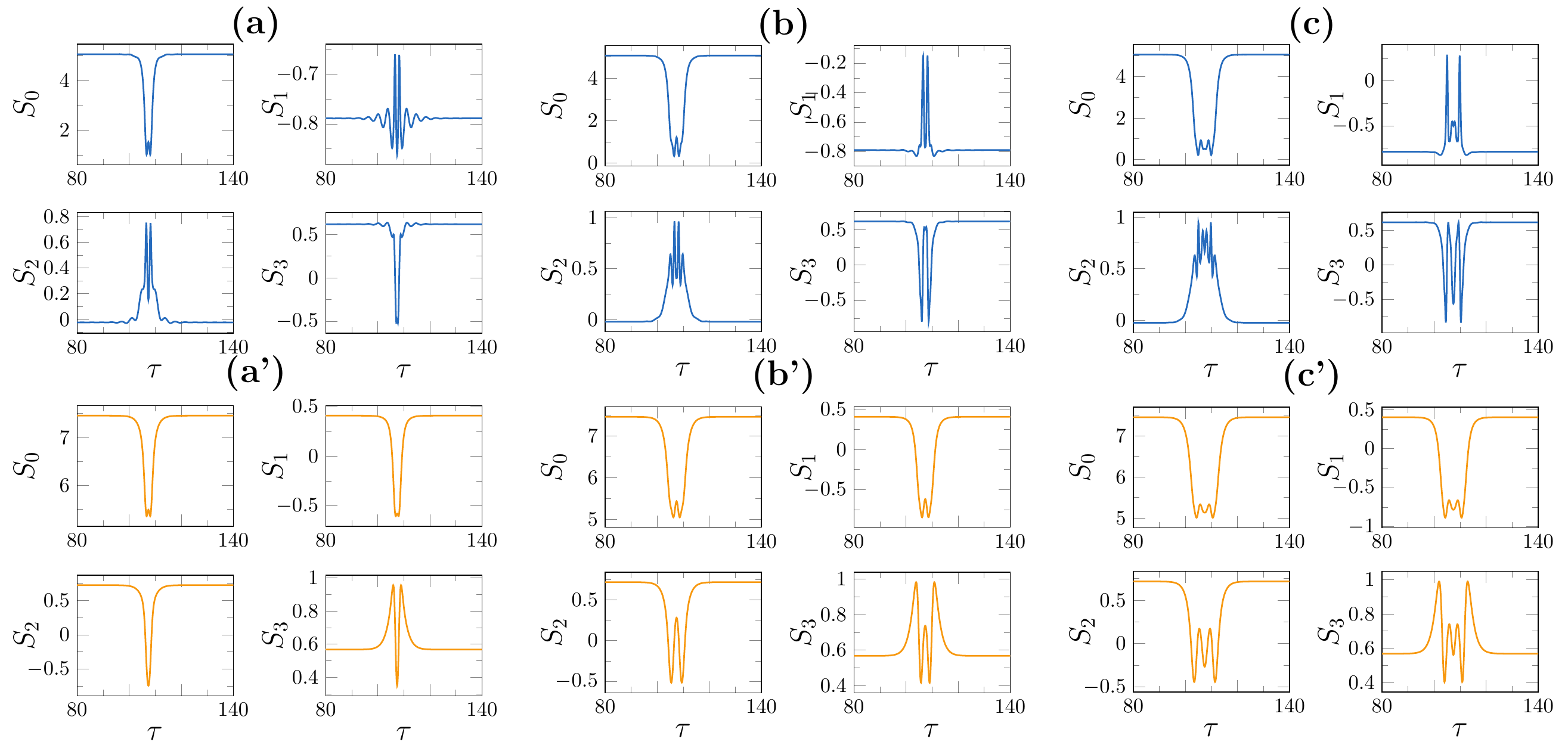}\\
	\caption{Profiles of the Stokes parameters $S_0$, $S_1$, $S_2$ and $S_3$ as a function of the fast time $\tau$ for the stable solutions indicated in Fig.~\ref{fig:tristab}. Parameters are the same.}
	\label{fig:tristabstokes}
\end{figure*}
\begin{figure}
	\includegraphics[width=\linewidth]{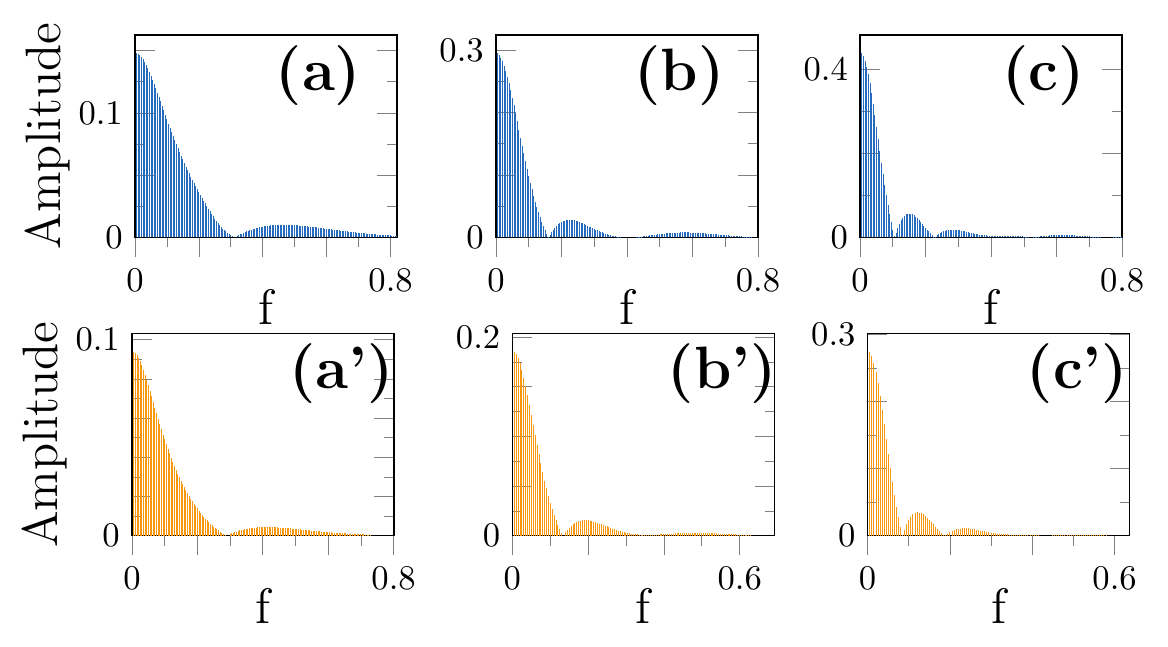}\\
	\caption{Vector Kerr combs corresponding to the stable solutions shown in Fig.~\ref{fig:tristabstokes}.}
	\label{fig:tristabprof}
\end{figure}
\begin{figure}
	\includegraphics[width=\linewidth]{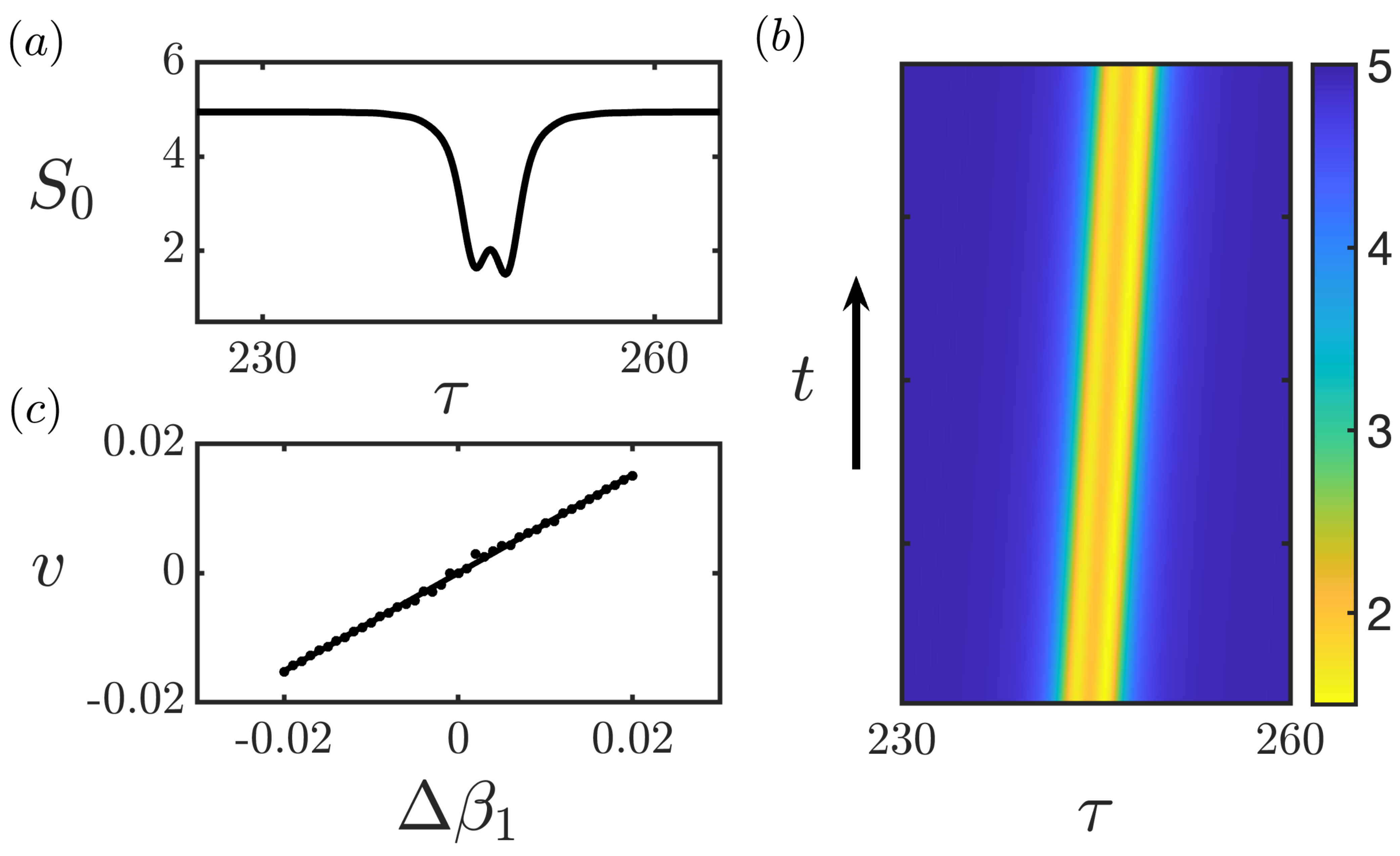}\\
	\caption{Moving vector TLS under the effect of GVM. (a) Profile of an asymmetric TLS due to non-zero GVM. (b) Space-time map of the same TLS showing the drifting motion. (c) Drift velocity $v$ as a function of the GVM coefficient $\Delta \beta_1$.}
	\label{fig:GVD}
\end{figure}

The profiles of the Stokes parameters corresponding to the solutions coexisting in the domain C are shown in Fig.~\ref{fig:tristabstokes}. Once again, they correspond to excursions for a small region of the cavity from a background corresponding to one of the high CW states into a lower CW state, this time corresponding to the states in Fig.~\ref{fig:LSA}(b) and Fig.~\ref{fig:HSSstokes}(b). As in the previous case, the excursions are between states with very different polarization properties, mostly regarding their ellipticity. The spectra of these solutions shown in Fig.~\ref{fig:tristabprof} are optical frequency combs again, exhibiting the same properties as in the previous case.

It has to be noted that we approximated in this work the group velocity mismatch to be zero $\Delta \beta_1 =0$, while it might not be the case in real conditions. The effect of a nonzero GVM is to introduce an asymmetry in the profiles shown in this work, and thus to create a drifting motion with constant velocity of the TLSs, as shown in Fig.~\ref{fig:GVD}. As can be seen from the figure, the speed is however so low that the drift it induces will be in most cases negligible compared to the cavity length. This effect will be studied more thoroughly in a future publication. 

\section{Conclusions}\label{sec:5}

We have presented a bifurcation analysis of temporal dissipative solitons and their corresponding frequency combs taking into account the polarization degree of freedom in microresonators subjected to continuous wave of linearly polarized injected fields. We have assumed that the optical resonator operates in a normal dispersion regime and far from any modulational instability. Thanks to the front locking mechanism, two different branches of temporal dissipative solitons can be generated in the system. We characterize these solutions by computing the Stokes parameters. Both branches of temporal dissipative solitons exhibit a collapse snaking type of bifurcation. We have considered first the double bistability with a well-separated hysteresis loop in which case the two branches do not exhibit an overlapping domain of stability. However, when the system develops a tristability, the two collapsed snaking overlap. We have shown that in this case the system develops a high degree of multistability of temporal dissipative solitons possessing different polarization states, different widths, and different power peaks in vectorial microresonators.\\

\begin{acknowledgments}
K.P. acknowledges the support by the Fonds Wetenschappelijk Onderzoek-Vlaanderen FWO (G0E5819N) and the Methusalem
Foundation.  M.T acknowledges financial support from the Fonds de la Recherche Scientifique FNRS under Grant CDR no. 35333527 "Semiconductor optical comb generator". A part of this work was supported by the "Laboratoire Associ\'{e} International" University of Lille - ULB on "Self-organisation of light and extreme events" (LAI-ALLURE)."
\end{acknowledgments}

\providecommand{\noopsort}[1]{}\providecommand{\singleletter}[1]{#1}%

\end{document}